\documentclass{aa}

\usepackage{graphicx}
\usepackage{color}
\usepackage[varg]{txfonts}
\bibpunct{(}{)}{;}{a}{}{,} 

\begin{document}

\title{Slipping reconnection in a solar flare observed in high resolution with the GREGOR solar telescope}

\author{M. Sobotka \inst{1} \and J. Dud\'ik \inst{1} \and C. Denker \inst{2} \and H. Balthasar \inst{2} \and
J. Jur\v{c}\'ak \inst{1} \and W. Liu \inst{1} \and T. Berkefeld \inst{3} \and M. Collados Vera \inst{4} \and
A. Feller \inst{5} \and A. Hofmann \inst{2} \and F. Kneer \inst{6} \and C. Kuckein \inst{2} \and
A. Lagg \inst{5} \and R. E. Louis \inst{2} \and O. von der L\"uhe \inst{3} \and H. Nicklas \inst{6} \and
R. Schlichenmaier \inst{3} \and \\ D. Schmidt \inst{7}\and W. Schmidt \inst{3} \and M. Sigwarth \inst{3} \and
S. K. Solanki \inst{5,8} \and D. Soltau \inst{3} \and J. Staude \inst{2} \and K. G. Strassmeier \inst{2} \and
R. Volkmer \inst{3} \and T. Waldmann \inst{3}}

 \institute{Astronomical Institute, Academy of Sciences of the Czech Republic, Fri\v{c}ova 298, 251 65 Ond\v{r}ejov, Czech Republic
 \and Leibniz Institute for Astrophysics Potsdam, An der Sternwarte 16, 14482 Potsdam, Germany
 \and Kiepenheuer Institute for Solar Physics, Sch\"{o}neckstra\ss e 6, 79104 Freiburg, Germany
 \and Instituto de Astrof\'isica de Canarias, V\'ia L\'actea, 38200 La Laguna (Tenerife), Spain
 \and Max Planck Institute for Solar System Research, Justus-von-Liebig-Weg 3, 37077 G\"ottingen, Germany
 \and Institute of Astrophysics, Georg-August University G\"ottingen, Friedrich-Hund-Platz 1, 37077 G\"ottingen, Germany
 \and National Solar Observatory, 3010 Coronal Loop, Sunspot, NM 88349, USA
 \and School of Space Research, Kyung Hee University, Yongin, Gyeonggi-Do, 446-701, Korea}

\date{Received December 15, 2015; accepted April 21, 2016}

\abstract
{A small flare ribbon above a sunspot umbra in active region 12205 was observed on November 7, 2014, at 12:00\,UT in the blue imaging channel of the 1.5 m GREGOR telescope, using a 1\,\AA ~\ion{Ca}{II} H interference filter. Context observations from the Atmospheric Imaging Assembly (AIA) onboard the Solar Dynamics Observatory (SDO), the Solar Optical Telescope (SOT) onboard Hinode, and the Interface Region Imaging Spectrograph (IRIS) show that this ribbon is part of a larger one that extends through the neighboring positive polarities and also participates in several other flares within the active region. We reconstructed a time series of 140 seconds of \ion{Ca}{II}~H images by means of the multiframe blind deconvolution method, which resulted in spatial and temporal resolutions of 0.1$\arcsec$ and 1\,s. Light curves and horizontal velocities of small-scale bright knots in the observed flare ribbon were measured. Some knots are stationary, but three move along the ribbon with speeds of 7--11 km s$^{-1}$. Two of them move in the opposite direction and exhibit highly correlated intensity changes, which provides evidence of a slipping reconnection at small spatial scales.}


\titlerunning{Slipping reconnection in a solar flare observed with GREGOR}
\authorrunning{M. Sobotka et al.}
\maketitle

\section{Introduction}
\label{Sect:1}

Solar flares are explosive phenomena that are characterized by a strong, rapid increase of electromagnetic radiation throughout the spectrum \citep[e.g.,][]{Fletcher11}. The flare energy originates in the stressed solar magnetic fields and is released by the process of magnetic reconnection \citep[e.g.,][]{Dungey53,Priest00,Zweibel09}. Newly reconnected magnetic field lines constitute various observed structures, such as flare loops and the erupting flux rope \citep[e.g.,][]{Dere99,Cheng13,Su13,Dudik14}. This process is described in two dimensions by the standard CSHKP solar flare model \citep[see e.g.,][]{Shibata11,Priest14}, which involves reconnection at a magnetic null-point.

However, in three dimensions (3D), the reconnection can also proceed in the absence of magnetic null-points and the associated topological discontinuities \citep{Longcope05}. In these cases, the reconnection occurs in the quasi-separatrix layers \citep[QSLs,][]{Priest95,Demoulin96a}, which are regions where the magnetic connectivity has strong gradients, but is still continuous. The QSLs thus still constitute boundaries between different flux systems. They are also associated with local enhancements of the electric current density \citep{Masson09,Wilmot09} and correspond well with the observed flare ribbons \citep{Demoulin97,Janvier14}. The magnetic reconnection in the QSLs is characterized by the apparent slipping motions of the field lines. This occurs because of the continuous exchange of connectivities between reconnecting flux systems \citep{Priest95,Priest03,Aulanier06,Aulanier12,Janvier13}. However, the slippage of the field lines may not correspond to the bulk plasma motions because the reconnection process is diffusive in nature.

%
   \begin{figure*}[t]
       \centering
        \includegraphics[width=6.45cm,clip,bb= 0 42 495 305]{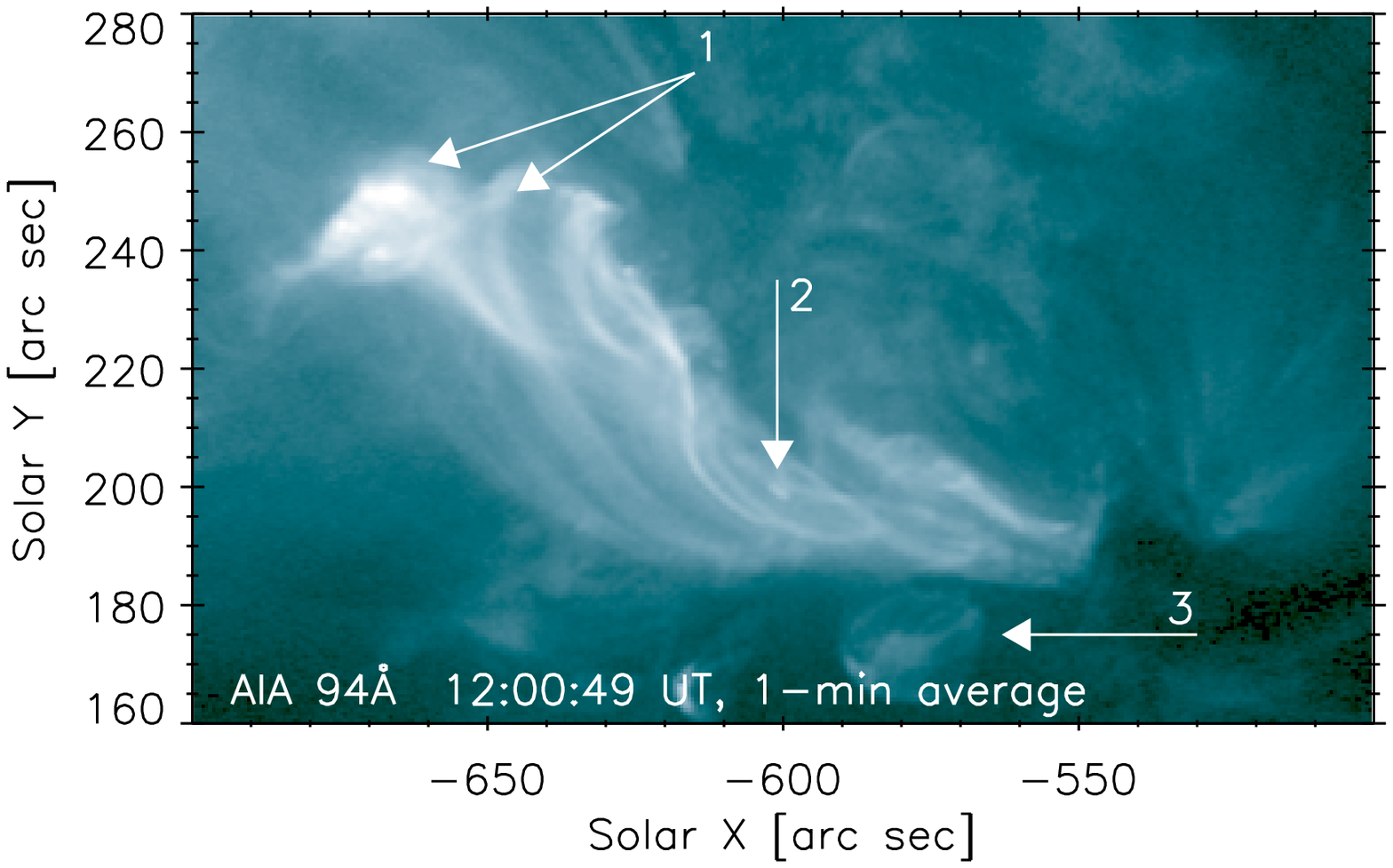}
        \includegraphics[width=5.58cm,clip,bb=67 42 495 305]{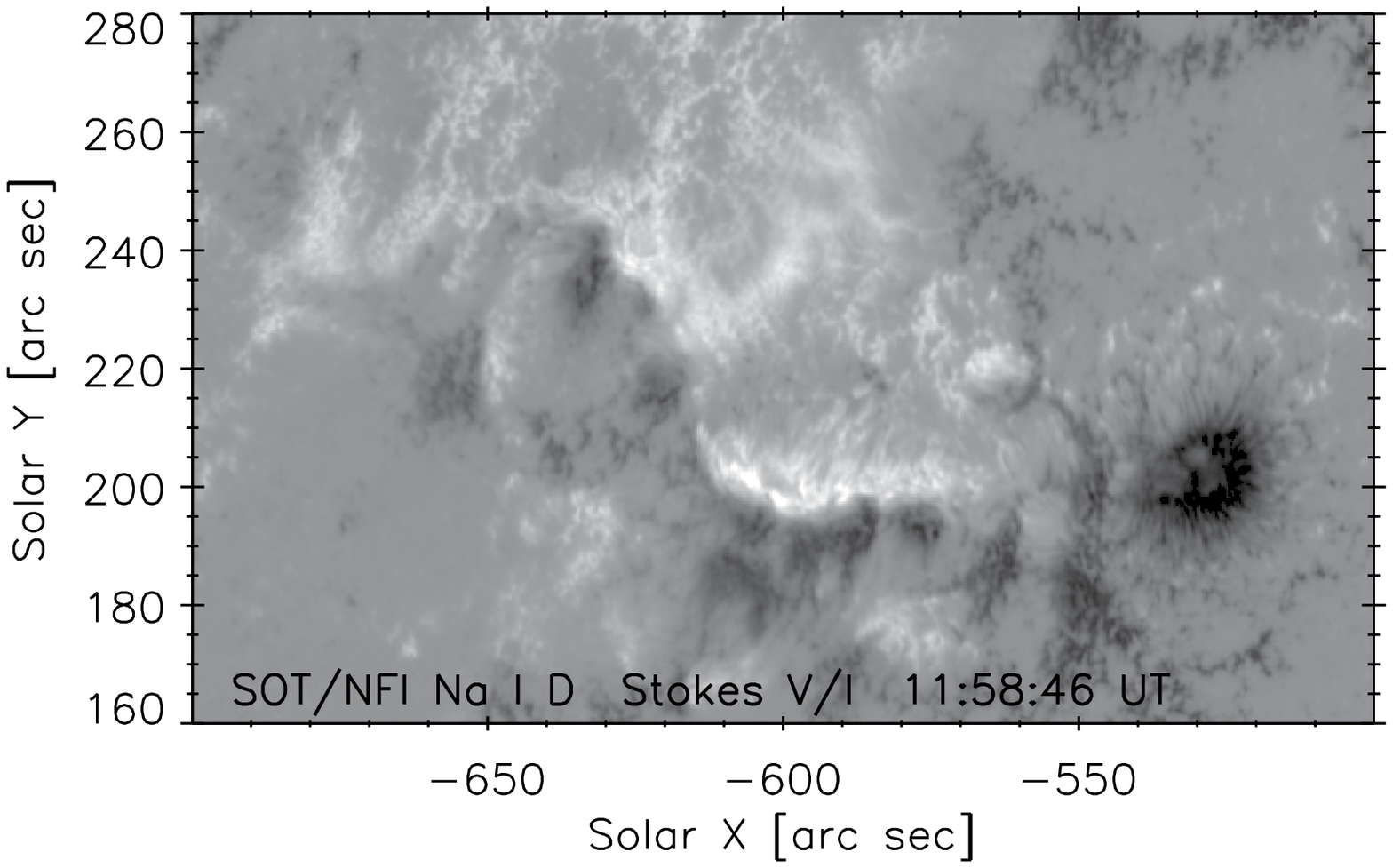}
        \includegraphics[width=5.58cm,clip,bb=67 42 495 305]{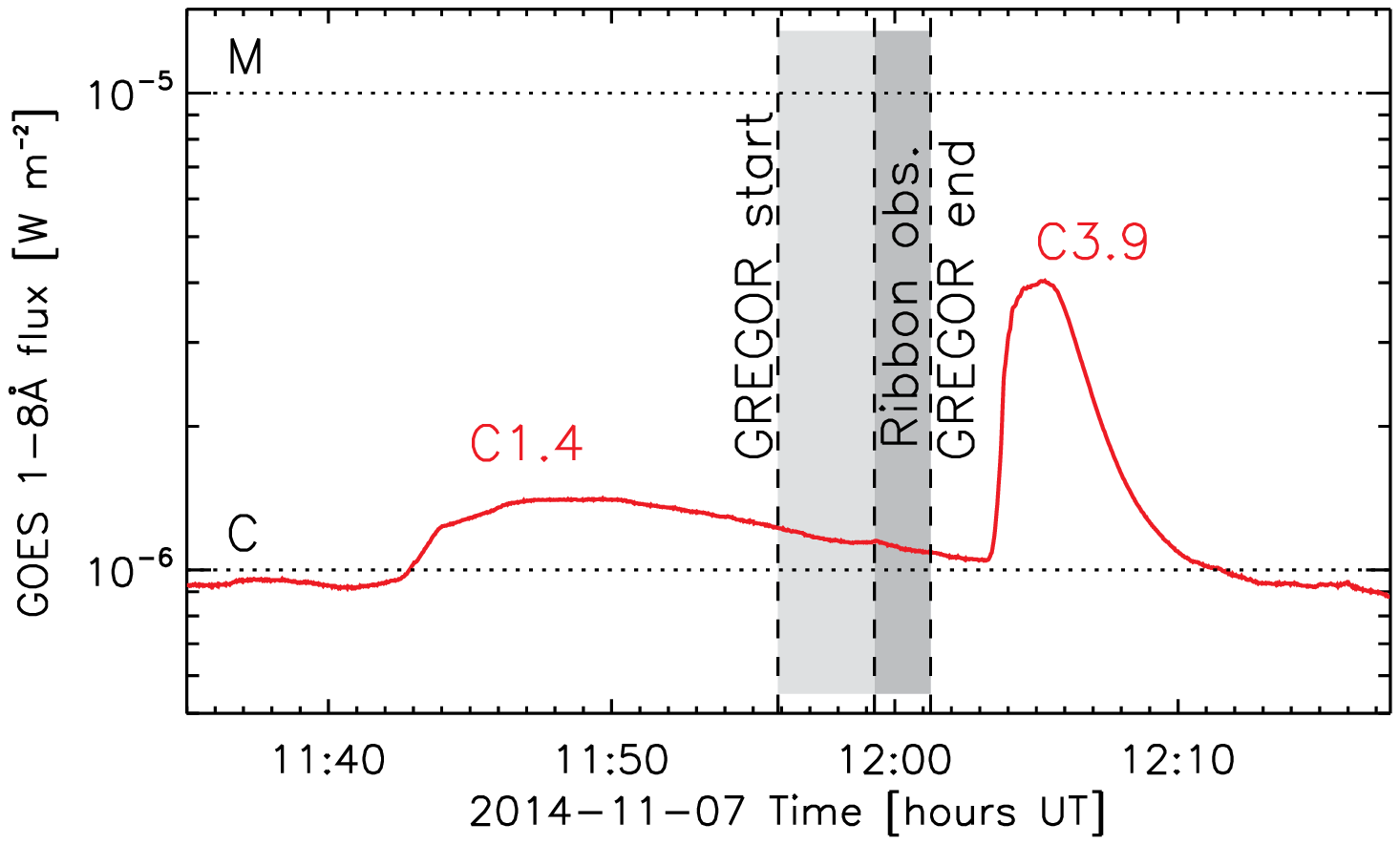}
        \includegraphics[width=6.45cm,clip,bb= 0 42 495 305]{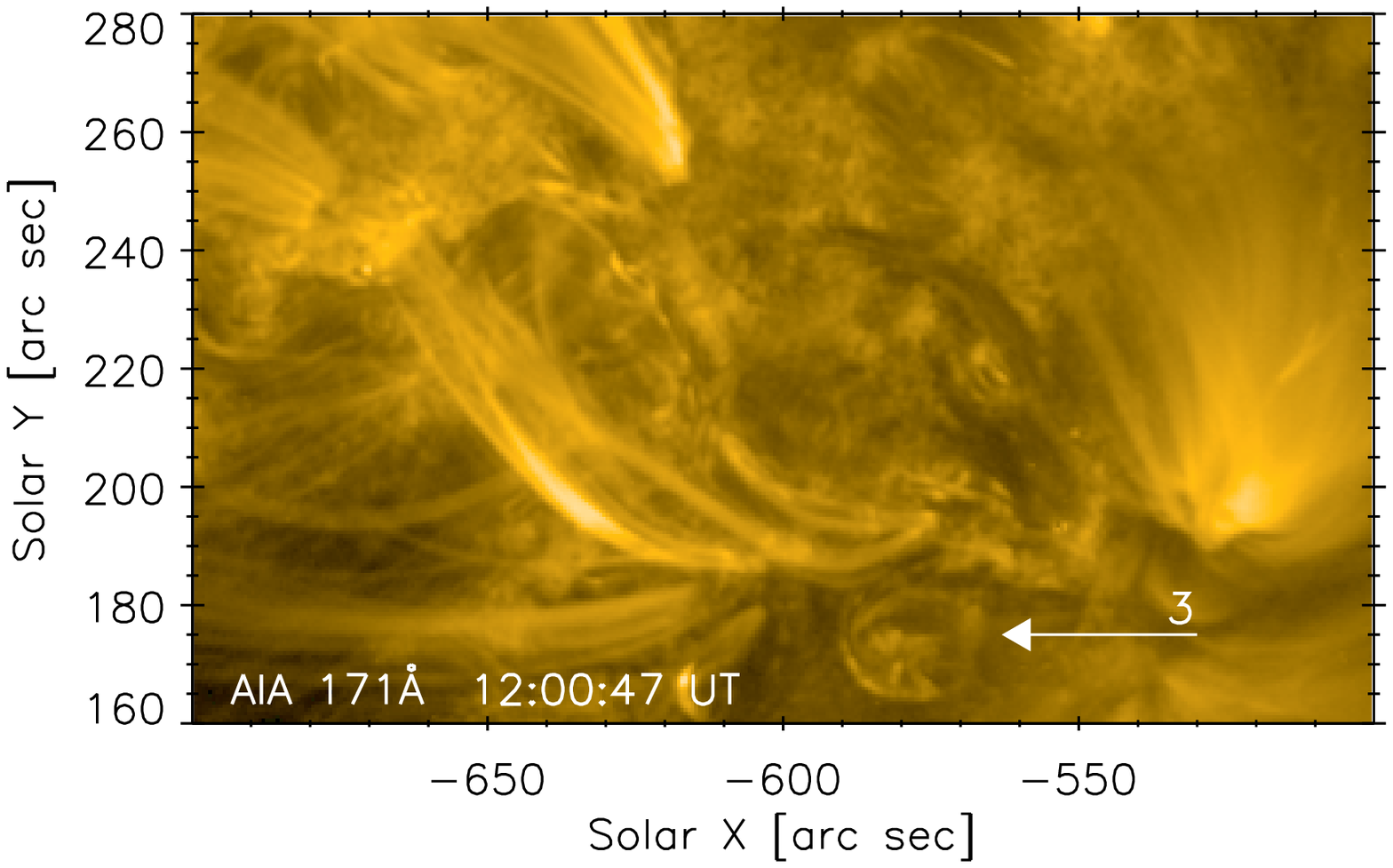}
        \includegraphics[width=5.58cm,clip,bb=67 42 495 305]{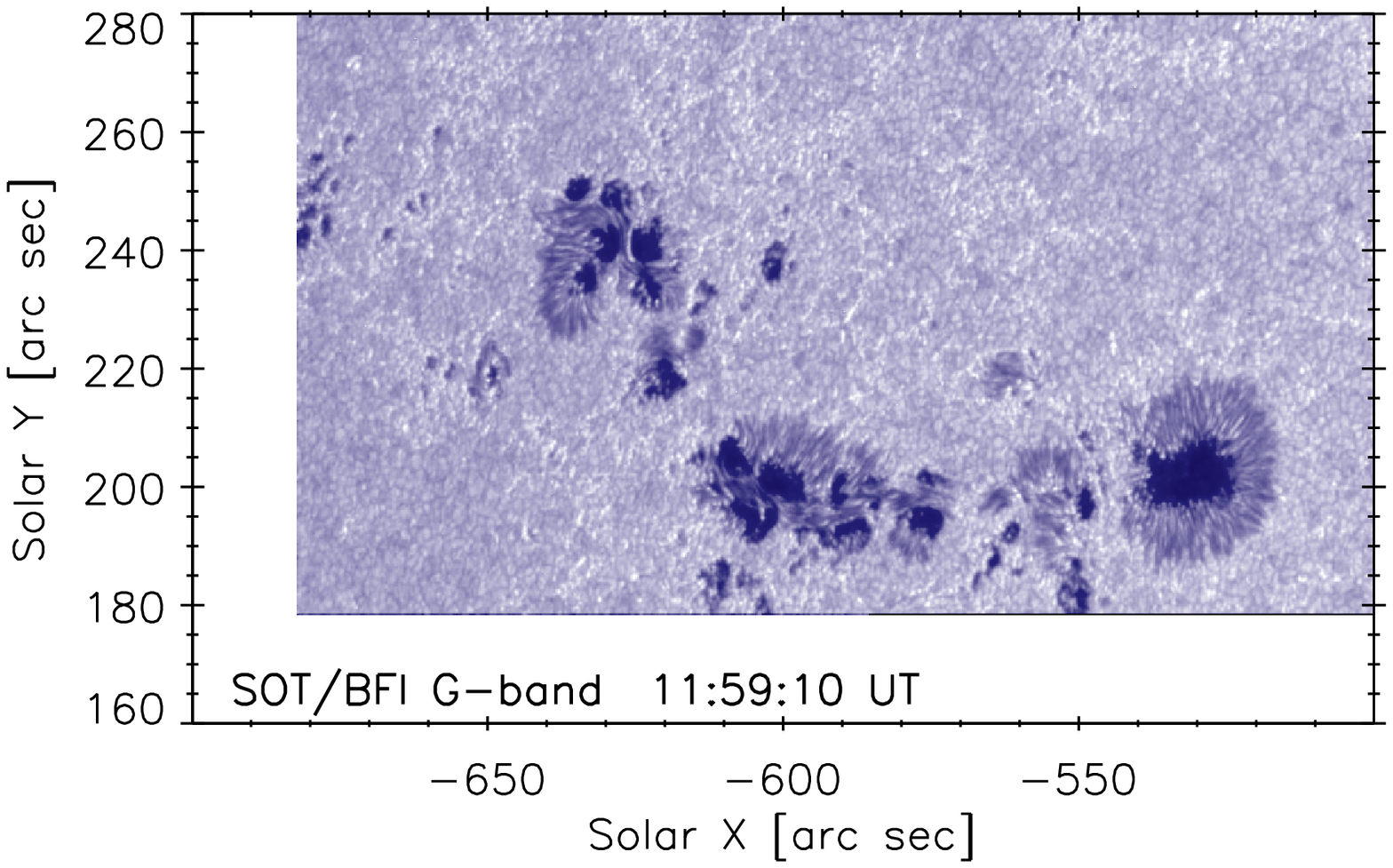}
        \includegraphics[width=5.58cm,clip,bb=67 42 495 305]{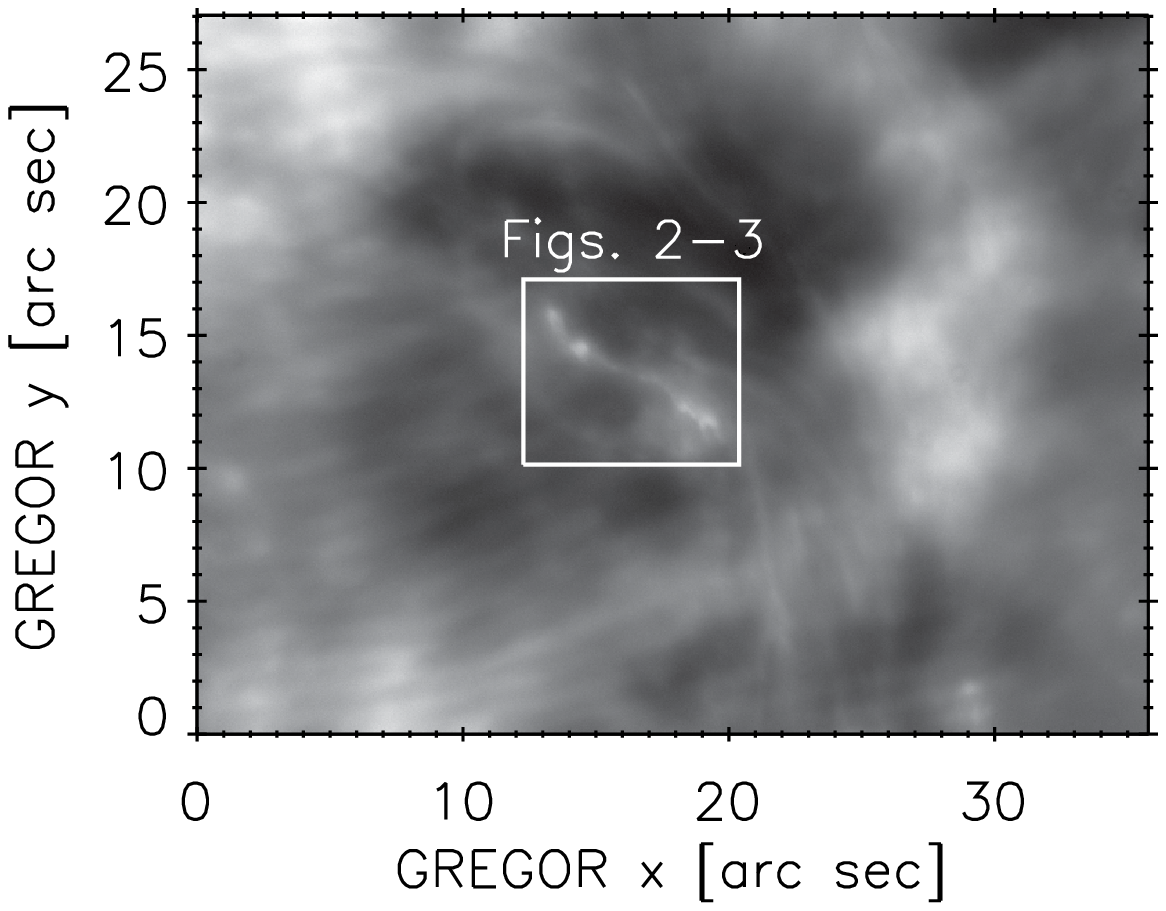}
        \includegraphics[width=6.45cm,clip,bb= 0  0 495 305]{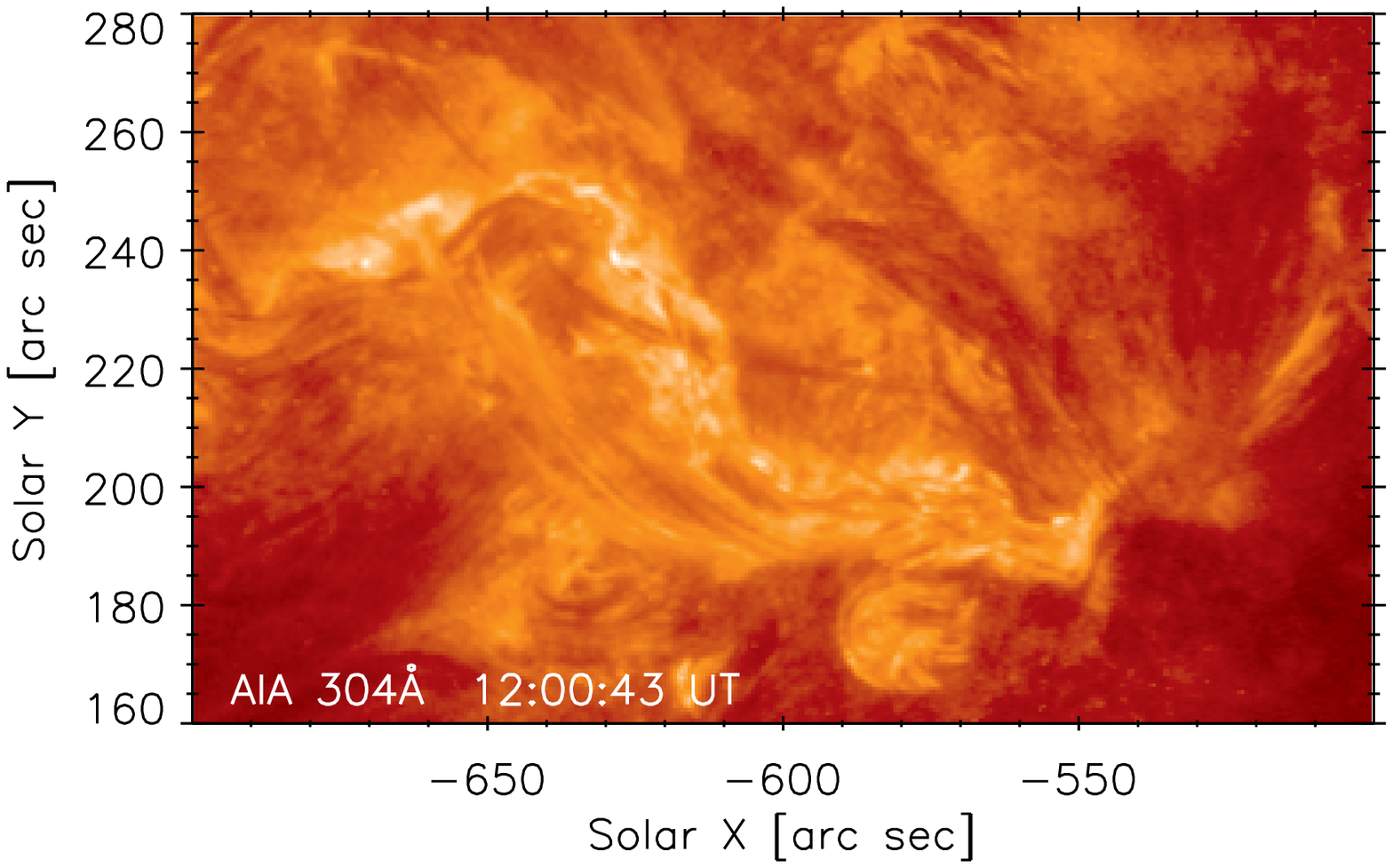}
        \includegraphics[width=5.58cm,clip,bb=67  0 495 305]{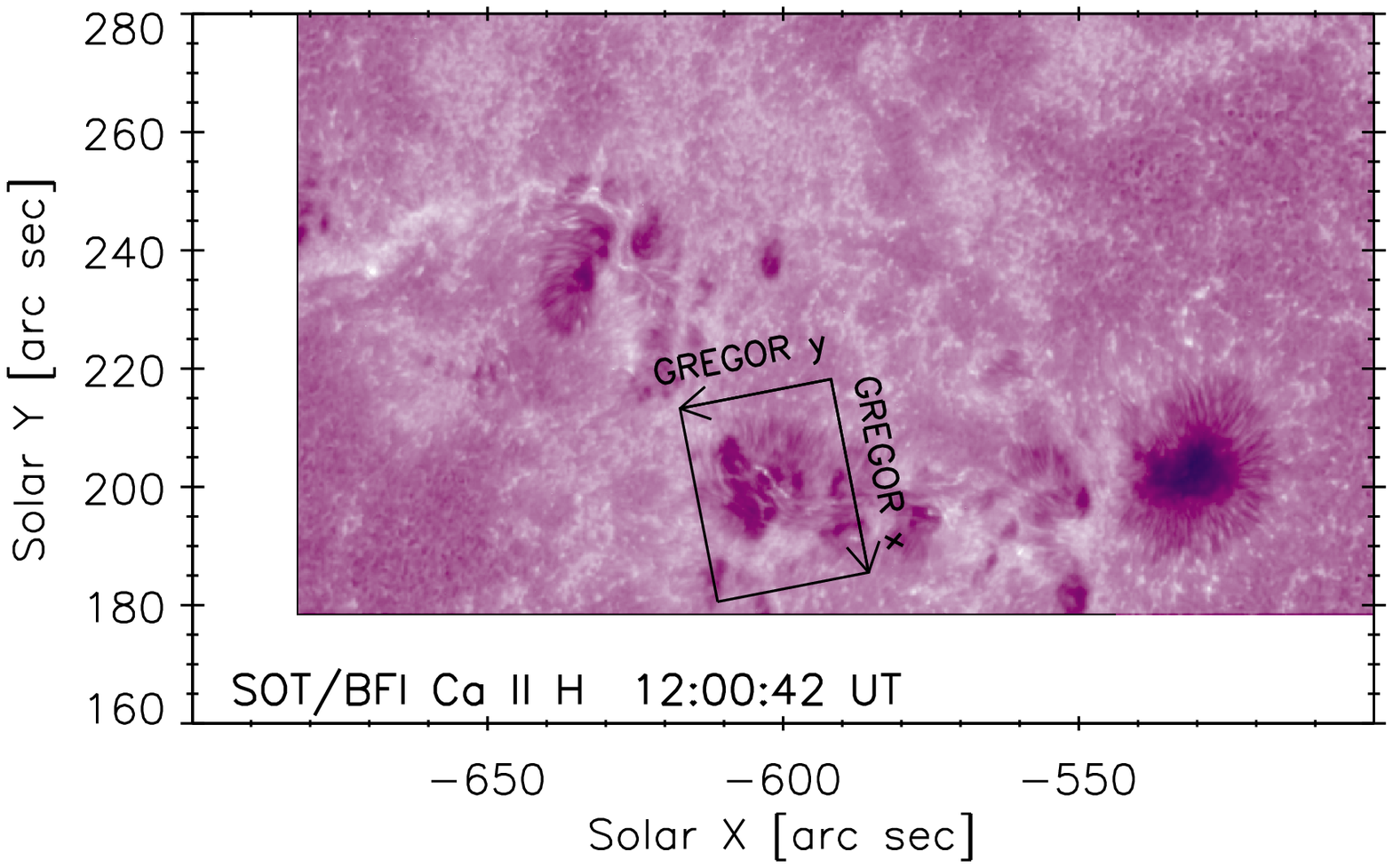}
        \includegraphics[width=5.58cm,clip,bb=67  0 495 305]{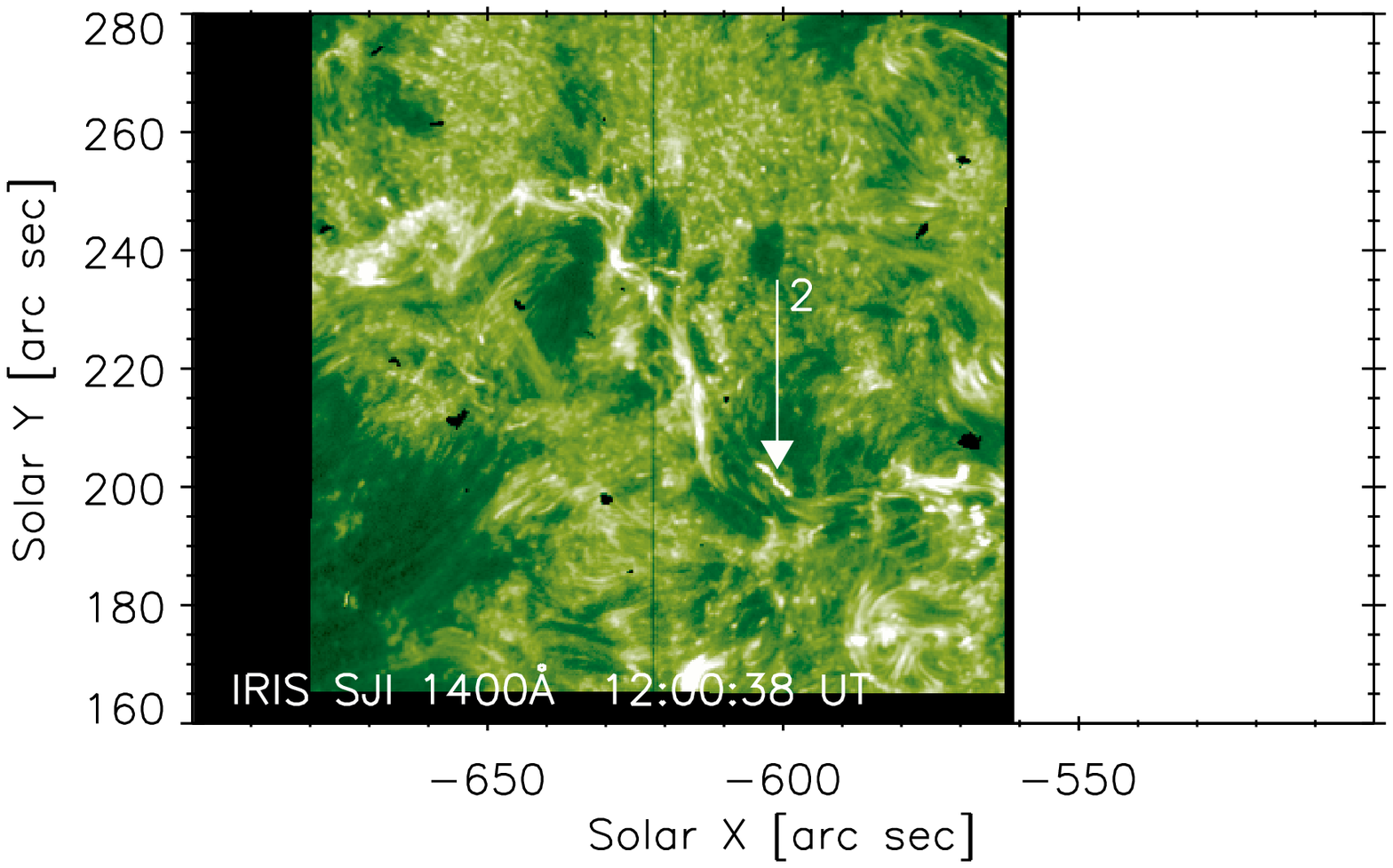}
        \caption{Context observations of AR 12205 from SDO/AIA (\textit{left}), Hinode/SOT (\textit{middle}), and the GOES X-ray flux, GREGOR \ion{Ca}{II} H, and IRIS 1400\,\AA~slit-jaw image (\textit{right}). Arrows 1--3 point to features discussed in the text. The GREGOR field of view, together with its axes, is shown as the black box in the \textit{bottom middle} panel. The GREGOR axes are rotated and flipped with respect to the solar $X$ and $Y$ coordinates. The white box corresponds to the field of view shown in Figs. \ref{Fig:Snapshots} and \ref{Fig:Tracking}. The temporal evolution of the AIA 94\,\AA~emission in the \textit{top left} panel is shown in a movie.}
       \label{Fig:Overview}
   \end{figure*}

Since QSLs are generalizations of the true topological discontinuities, slipping reconnection that occurs within them is a more general mode of magnetic reconnection. This means that it should be widely observed, especially in solar flares. However, reports of slipping reconnection were very rare \citep{Aulanier07,Testa13} until very recently, when this reconnection was found in several solar flares \citep{Dudik14,Li14,Li15}. The first of these reports, based on data from the 131~\AA~channel of the Atmospheric Imaging Assembly \citep[AIA,][]{Lemen12,Boerner12} onboard the Solar Dynamics Observatory \citep[SDO, ][]{Pesnell12} pertained to the early phase of a long-duration X-class flare on July 12, 2012 \citep{Dudik14}. The apparent slipping motion of flare loops was clearly distinguishable from the flare onset and was not obscured by line-of-sight effects in optically thin plasma. Using AIA data, \citet{Li14,Li15} reported the apparent slipping motion in other flares, including in their impulsive phases. Furthermore, \citet{Li15} found quasi-periodic patterns in small-scale bright knots that moved along the ribbon observed by the Interface Region Imaging Spectrograph spacecraft \citep[IRIS, ][]{DePontieu14}. These observations of the slipping reconnection verify the predictions of the generalized 3D reconnection models \citep{Aulanier06,Aulanier12,Janvier13} and signify that the energy release in solar flares is indeed an intrinsically 3D phenomenon.

In this paper, we report on optical imaging observations with ultra-high spatial and temporal resolution performed in the blue imaging channel \citep[BIC,][]{Puschmann12} in the line core of \ion{Ca}{II} H 3968\,\AA~ with the new 1.5 meter GREGOR telescope \citep{Schmidt12} at the Observatorio del Teide, Spain. Small-scale slipping knots are observed in a flare ribbon located in a $\delta$-type sunspot. Compared to the previous space-borne observations of slipping motion, our data have spatial and temporal resolutions higher by a factor of 3 and 10, respectively. Moreover, they show for the first time that the apparent slipping motion can also be detected in chromospheric line emission. Context observations from SDO/AIA, IRIS, and the Solar Optical Telescope \citep[SOT,][]{Tsuneta08} onboard Hinode are described in Sect.~\ref{Sect:2}. GREGOR observations and a kinematical analysis of the slipping knots are reported in Sect.~\ref{Sect:3}. The results are discussed in terms of slipping reconnection models in Sect.~\ref{Sect:4}, and conclusions are drawn in Sect.~\ref{Sect:5}.

\section{Multiwavelength context observations}
\label{Sect:2}

Active region NOAA 12205 (hereafter, AR 12205) was located at position 13N, 37E on November 7, 2014. This active region had a $\beta\gamma\delta - \beta\gamma\delta$ magnetic configuration and was the site of many flares, including six C-class flares, two M-class flares, and one X1.6-class flare on November 7 alone. The active region is displayed in Fig.~\ref{Fig:Overview} as observed at multiple wavelengths by different space-borne instruments. This figure shows the active region at approximately 12:00 UT, corresponding to the time of the observations performed by the GREGOR telescope, which ran during the period 11:55:13--12:01:31\,UT (Sect.~\ref{Sect:3}).

Shortly after the GREGOR observations, the GOES {1--8\,\AA} X-ray flux began to rise at 12:04\,UT and peaked at 12:06\,UT during a C3.9 flare.
The site of the C3.9 flare was a circular ribbon indicated by arrow 3 in Fig.~\ref{Fig:Overview}. Its location and several transient brightenings preceding the flare were visible already around 12:00\,UT. These transient brightenings occurred mostly along a long, ribbon-like structure that is apparent in the Hinode/SOT \ion{Ca}{II} H observations, in the IRIS 1400\,\AA~slit-jaw images, or in the SDO/AIA observations in the 304\,\AA~filter. However, in some passbands the ribbon-like emission was overlaid with bright threads in the neighboring filament that lay along the magnetic inversion line (Fig.~\ref{Fig:Overview}). This ribbon-like structure corresponded to the footpoints of hot \ion{Fe}{XVIII} loops observed in the AIA 94\,\AA~band that
were located in the positive magnetic polarities (Fig.~\ref{Fig:Overview}). The magnetogram presented in Fig.~\ref{Fig:Overview} shows the Stokes $V/I$ intensity in the red wing of the \ion{Na}{I} D-line as observed with the Hinode/SOT narrowband filter imager.

%
   \begin{figure*}[!ht]
       \centering
\includegraphics[width=17cm]{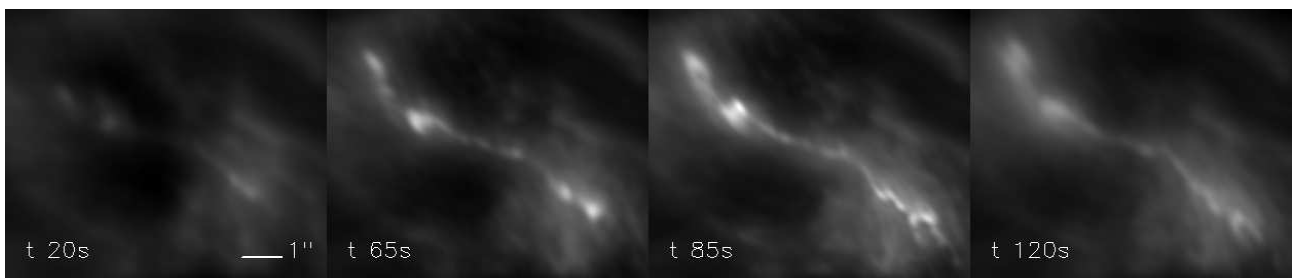}
\caption{GREGOR snapshots of the flare ribbon at 11:59:08 UT + $t$ restored with the multiframe blind deconvolution. The field of view 8.1\arcsec$\times$7\,\arcsec \,  corresponds to the white box shown in the \textit{middle right} panel of Fig.~\ref{Fig:Overview}. The temporal evolution is shown in a movie.}
       \label{Fig:Snapshots}
   \end{figure*}
%
   \begin{figure*}[]
       \centering
\includegraphics[width=8.8cm]{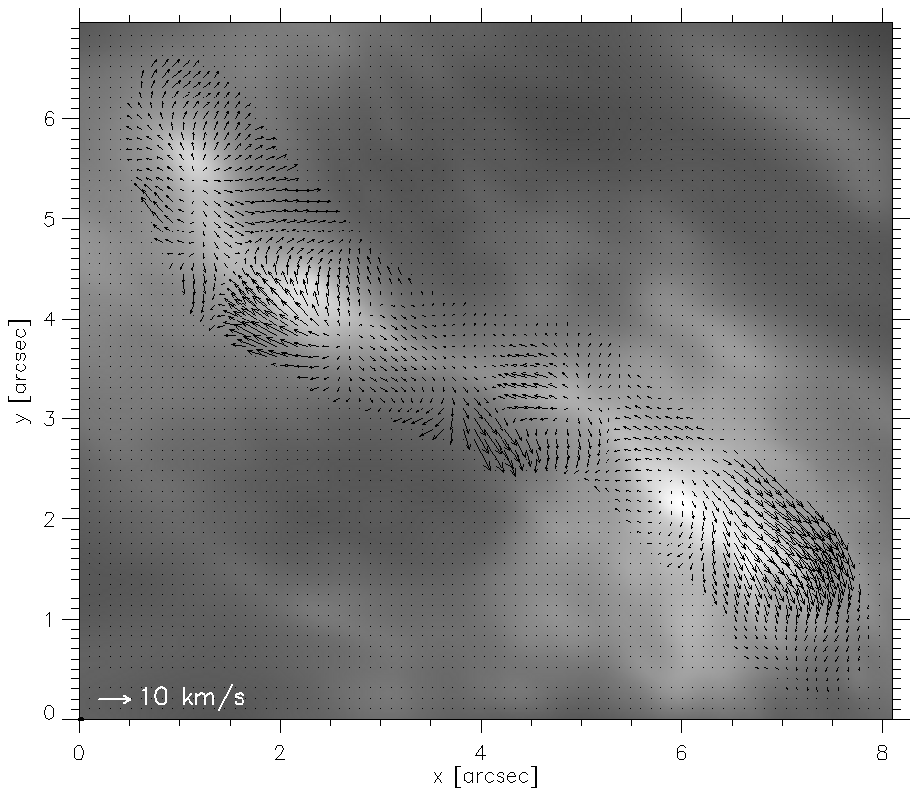}
\includegraphics[width=8.8cm]{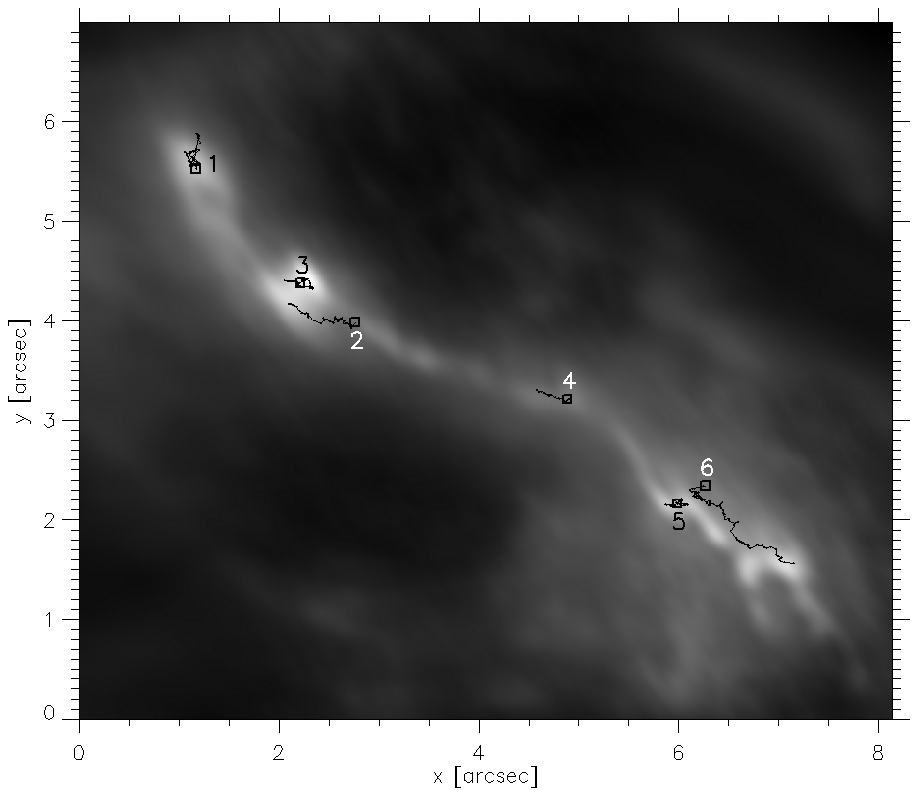}
\caption{Motion tracking in the flare ribbon. \textit{Left:} LCT flow map with an underlying average image of the series. \textit{Right:} Trajectories of knots 1--6 with the best image in the background. Small squares denote starting positions of the tracking. The field of view shown is the same as in Fig.~\ref{Fig:Snapshots}.}
       \label{Fig:Tracking}
   \end{figure*}

The hot \ion{Fe}{XVIII} loops are seen to be evolving in the attached movie 1, with a particular flare-like brightening in the northern part that is visible from 11:39\,UT onward and shows signatures of slipping reconnection at the location of arrow 1 until at least 11:50\,UT (see Fig.~\ref{Fig:Overview} and movie~1). The maximum occurred at 11:47 as a weak C1.4-class event (Fig. \ref{Fig:Overview} \textit{top right}). These brightenings are followed by several other brightenings that occurred successively southward along the ribbon and reached the position indicated by arrow 2 at about 12:00 UT. This is the portion of the ribbon within the $\delta$-spot observed by GREGOR (Sect.~\ref{Sect:3}). The brightenings are finally followed by the C3.9 flare at the location of the circular ribbon (arrow 3). The portion of the ribbon indicated by arrow 2 appears to be disconnected from the main ribbon. However, the long ribbon, the sunspot-portion of the ribbon, and the central bright part of the circular ribbon are all located in the positive magnetic polarities. The conjugate footpoints of the flare loops, as seen in AIA 94\,\AA, are located in the negative polarities in the vicinity of the leading sunspot (Fig.~\ref{Fig:Overview}).

The coalignment between different observations presented in Fig.~\ref{Fig:Overview} was achieved manually using a multistep process that is based on successive coalignments between datasets in which similar morphological structures are observed. The AIA 1600\,\AA~data (not shown) were used to determine the solar $X$ and $Y$ coordinates. The Hinode/SOT \ion{Ca}{II} H and IRIS 1400\,\AA~slit-jaw observations were coaligned with 1600\,\AA. Other SOT observations were coaligned with the SOT \ion{Ca}{II} H, as were the GREGOR \ion{Ca}{II} H observations. We note that the GREGOR observations are rotated and flipped with respect to the other observatories, see Fig.~\ref{Fig:Overview}, \textit{bottom middle} panel.

%
   \begin{figure*}[]
       \centering
\includegraphics[width=8.8cm]{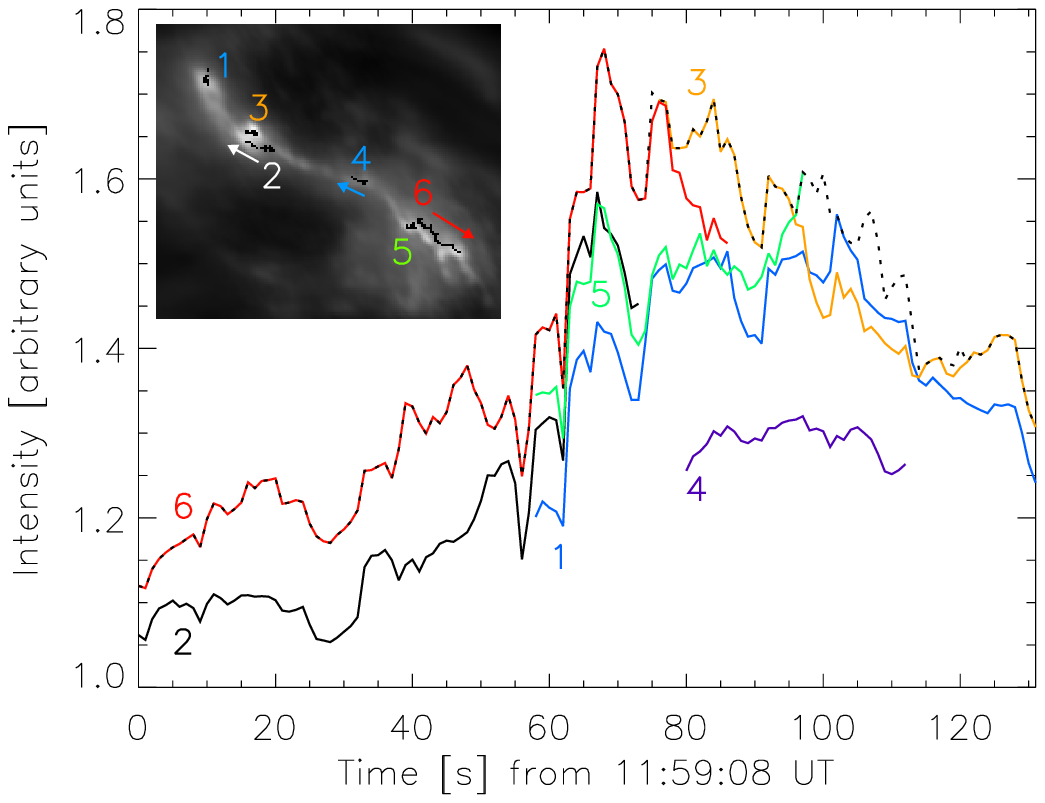}
\includegraphics[width=8.8cm]{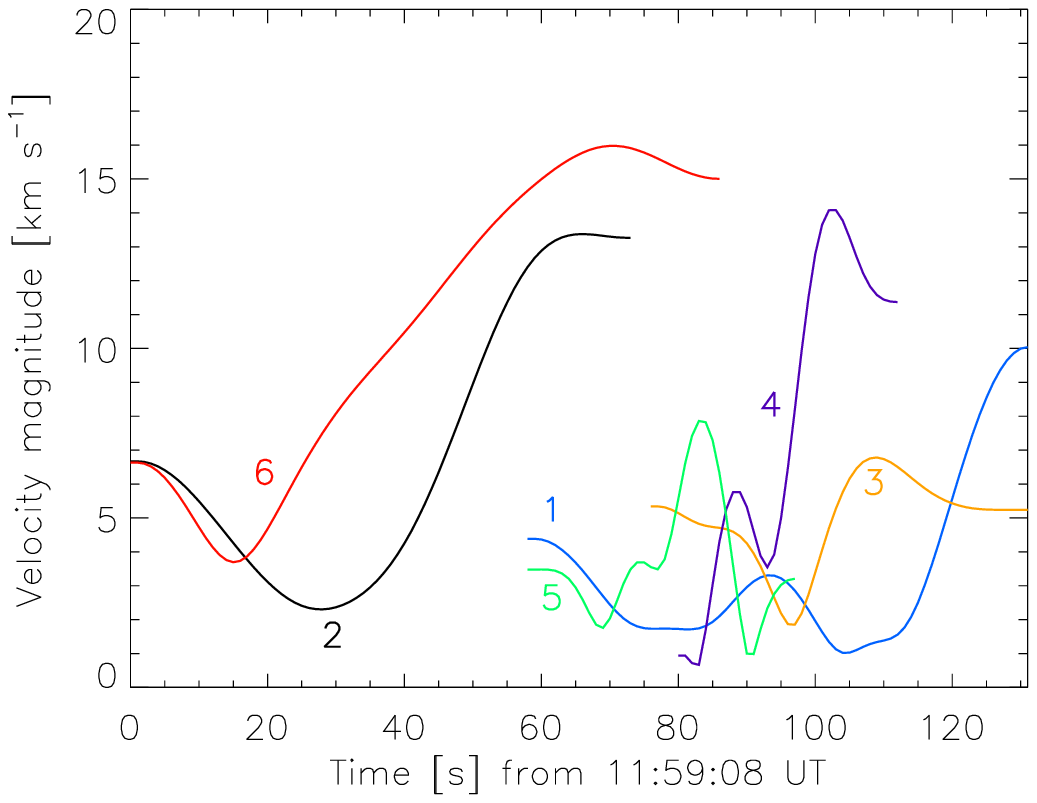}
\caption{Intensities and horizontal velocities of bright knots 1--6. \textit{Left:} Light curves. The intensities are given in units of the initial intensity before the flare. The black dotted line represents the absolute intensity maximum in the ribbon. \textit{Right:} Low-pass-filtered instantaneous velocity magnitudes.}
       \label{Fig:Curves}
   \end{figure*}

\section{Slipping knots observed by GREGOR}
\label{Sect:3}

The GREGOR observations were made during the early-science period of the telescope, using the integrated adaptive optics system \citep{Berkefeld12}. The field of 35\arcsec$\times$\,27\arcsec, including the ribbon and the underlying sunspot, was observed in the blue imaging channel of the GREGOR Fabry-P\'erot Interferometer \citep[GFPI,][]{Puschmann12}, which is equipped with a 1\,\AA~\ion{Ca}{II}\,H interference filter centered on 3968\,\AA. A series of \mbox{12-bit} frames was acquired with a frame rate of 7.677\,Hz, an exposure time of 30\,ms, and an image scale of 0.026$\arcsec$ per pixel from 11:55:13 to 12:01:31 UT. After the standard dark- and flat-field corrections, some residual fringes specific of the pointing position of the telescope remained in the field of view. To remove them, we averaged the frames at the beginning of the series (11:55:13--11:59:07 UT, light-shaded in Fig. \ref{Fig:Overview} \textit{top right}), that is, before the ribbon appeared. The rest of the series with the ribbon was divided by this average.
In addition to the fringes, this operation also removed real structures that were present before the ribbon
appeared, including the sunspot. The ribbon was not affected,
and the resulting series of images (\mbox{11:59:08--12:01:31 UT}, marked ``ribbon obs.'' in Fig. \ref{Fig:Overview} \textit{top right}) had intensities normalized to those before the flare beginning. For simplicity, we refer to the ribbon observed by GREGOR as a ``flare'', even though it is not a separate GOES X-ray event, but only a portion of a complex flaring activity, as discussed in Sect.~\ref{Sect:2}.

The multiframe blind deconvolution method \citep[MFBD,][]{Lofdahl02,Noort05,delaCruz15} was then applied to this series. Sequences of 32 frames were used to obtain one restored image, and the restoration was made in a sliding mode, with a step of 8 frames (frames 1--32 produced the first restored image, frames 9--40 the second, etc.), so that 132 restored images with a time step of 1.04\,s were obtained. Taking the smallest resolved features into account, the resulting spatial resolution was 0.1$\arcsec$. An example of four snapshots of the restored series is displayed in Fig.~\ref{Fig:Snapshots}, and the full restored series is shown in movie 2.

Small bright knots are clearly resolved in the ribbon. The local correlation tracking algorithm \citep[LCT, ][]{NovSim88} with a tracking window of 0.5$\arcsec$ was applied to the whole restored series to estimate the dynamics of the ribbon (Fig.~\ref{Fig:Tracking}, \textit{left}). In addition to the expansion of the ribbon area, motions with speeds of 5--12~km\,s$^{-1}$ in both directions along the ribbon are detected in three places. These velocities are generally lower than those reported from other flares \citep{Dudik14,Li14,Li15}, but we note that the slipping velocity may depend on the flare phase \citep{Dudik14}, and the observational reports tend to favor the impulsive phases of larger flares \citep{Dudik14,Li14,Li15}.

More information can be obtained by tracking individual bright knots. The procedure locks on the tracked feature using the spatial correlation of intensities between two neighboring frames in a 1.24$\arcsec$ $\times$ 1.24$\arcsec$ window around the feature and records the value and $(x,y)$ position of its intensity maximum in each frame. Six bright knots (1--6) were tracked successfully, and their trajectories are shown in the \textit{right} panel of Fig.~\ref{Fig:Tracking}. As a result of the splitting and merging effects that interrupted the tracking, they were not followed during their whole lifetime. The tracking periods range from 32 s (knot 4) to 86 s (knot 6).

The resulting light curves and velocity profiles of the tracked knots are shown in Fig.~\ref{Fig:Curves}. The intensities are given in units of the initial intensity before the flare. The inset in Fig.~\ref{Fig:Curves} \textit{left} shows the positions of the knots and the orientations of their motions. Knots 2 and 6, tracked from the beginning of the series, first show a slow increase in brightness, then a sudden brightening ($t = 63$~s after the beginning, seen also in knots 1 and 5), an intensity maximum (knot 6, $I = 1.75$, $t = 68$ s), and a decrease. The linear correlation coefficient for the intensity evolution of knots 2 and 6 during their common 73\,s tracking time is high, 0.97. This suggests that a common process is at the origin of both knots.

Three knots (knots 1, 3, and 5) follow a similar intensity evolution, but their positions oscillate randomly around the initial positions. Knot 4 appears later in time in the middle of the ribbon. Although its life and trajectory are short, it clearly translates in one direction (Fig.~\ref{Fig:Tracking}), in contrast to knots 1, 3, and 5. Knots 2 and 4 show a translatory motion along the ribbon in one direction, while knot 6 moves in the opposite direction. The instantaneous velocities of the knots calculated from their positions are very noisy. A shift of one pixel represents 18.75~km\,s$^{-1}$. By applying a low-pass Fourier filter, we reduced the error to $\pm 4$ km\,s$^{-1}$. The resulting time-dependent velocity magnitudes are plotted in the \textit{right} panel of Fig.~\ref{Fig:Curves}, showing acceleration phases of 0.2--0.7~km\,s$^{-2}$ that are typical for the moving knots 2, 4, and 6. These knots have average velocities (and standard deviations) of $7 \pm 1$, $8 \pm 2$, and $11 \pm 2$~km\,s$^{-1}$, respectively. The opposite motion of knots 2 and 6, together with their highly correlated intensity evolution, indicates the slipping reconnection.

%
   \begin{figure*}
       \centering
        \includegraphics[width=6.45cm,clip,bb= 0 42 495 370]{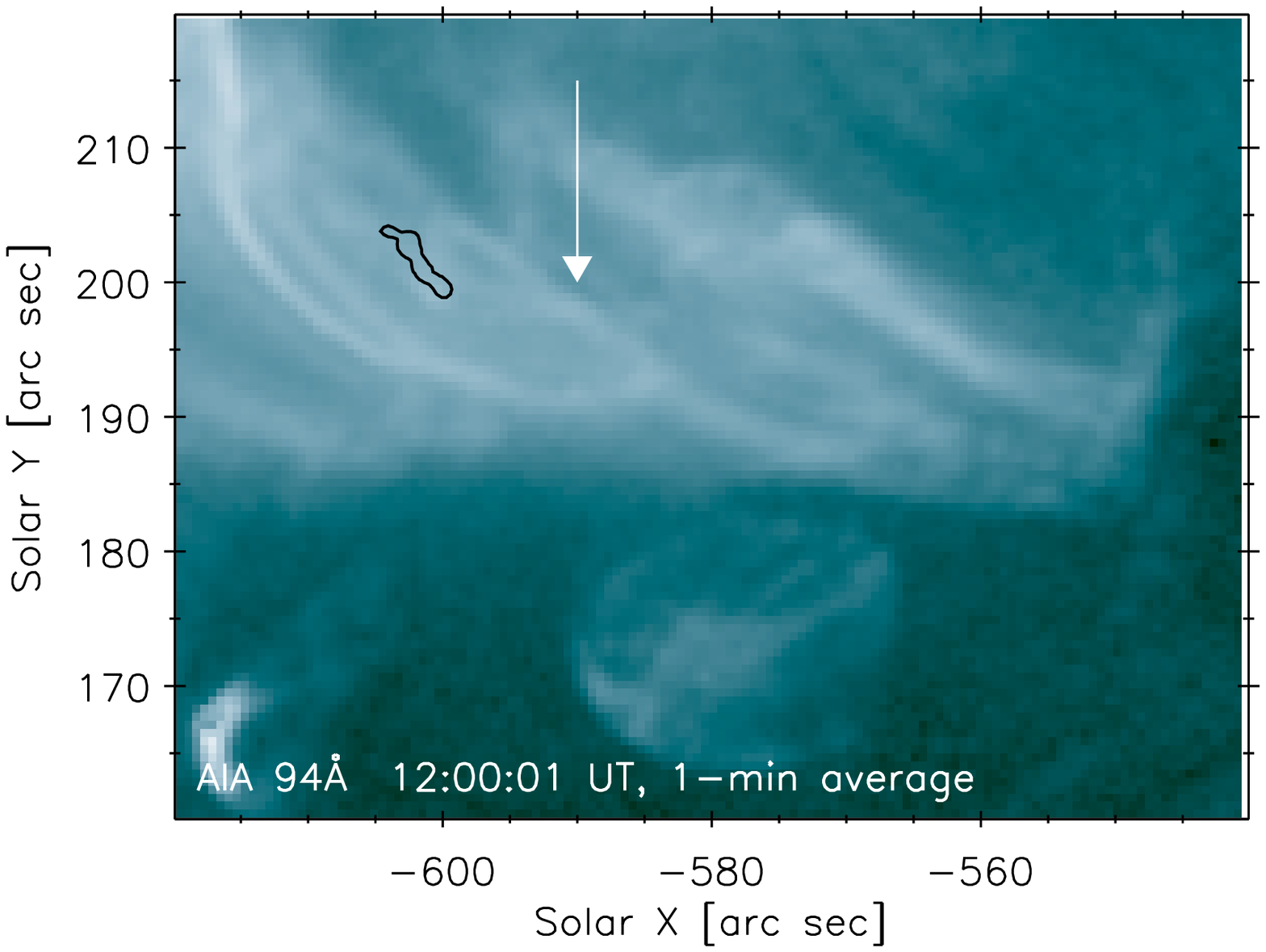}
        \includegraphics[width=5.58cm,clip,bb=67 42 495 370]{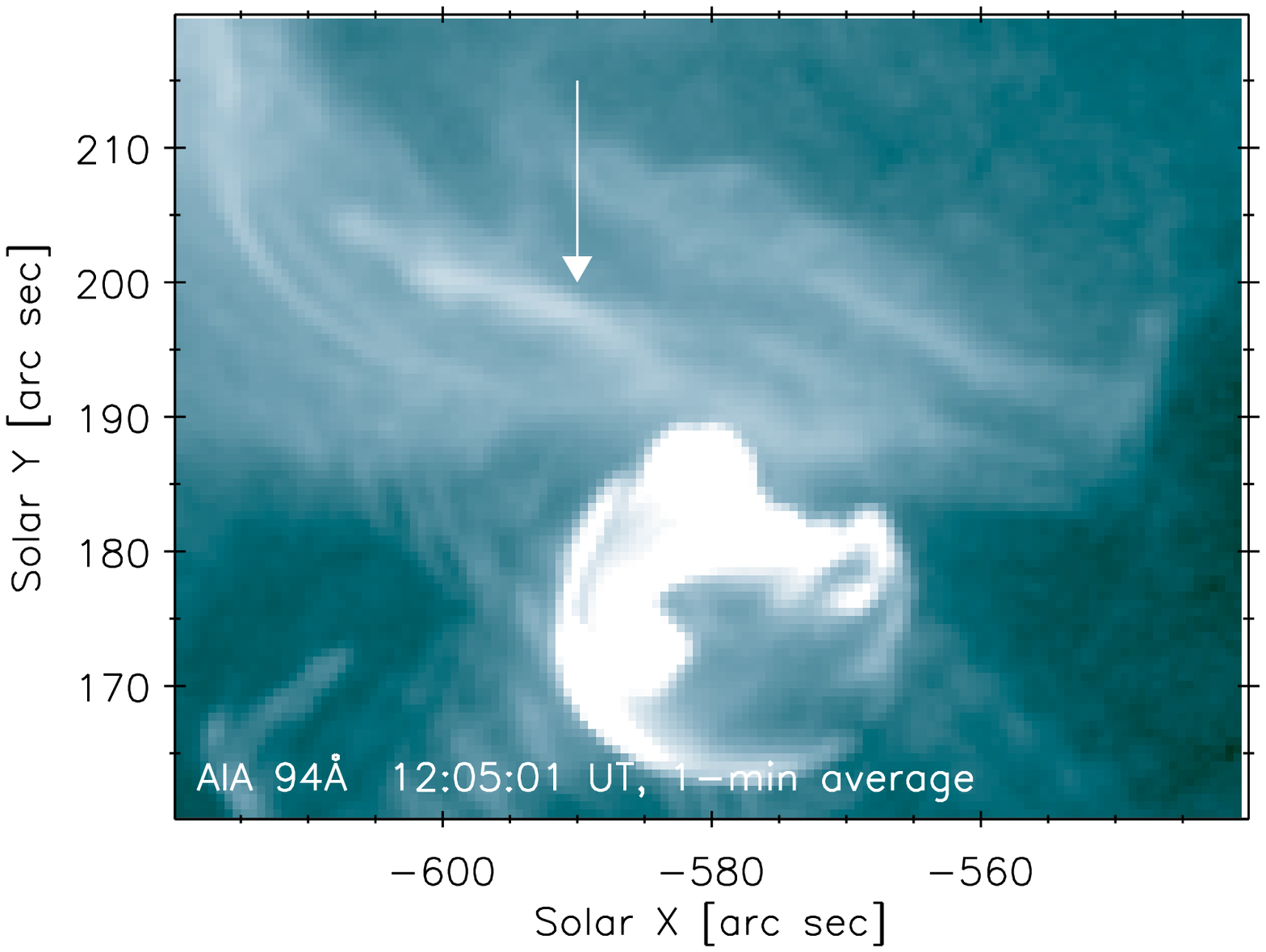}
        \includegraphics[width=5.58cm,clip,bb=67 42 495 370]{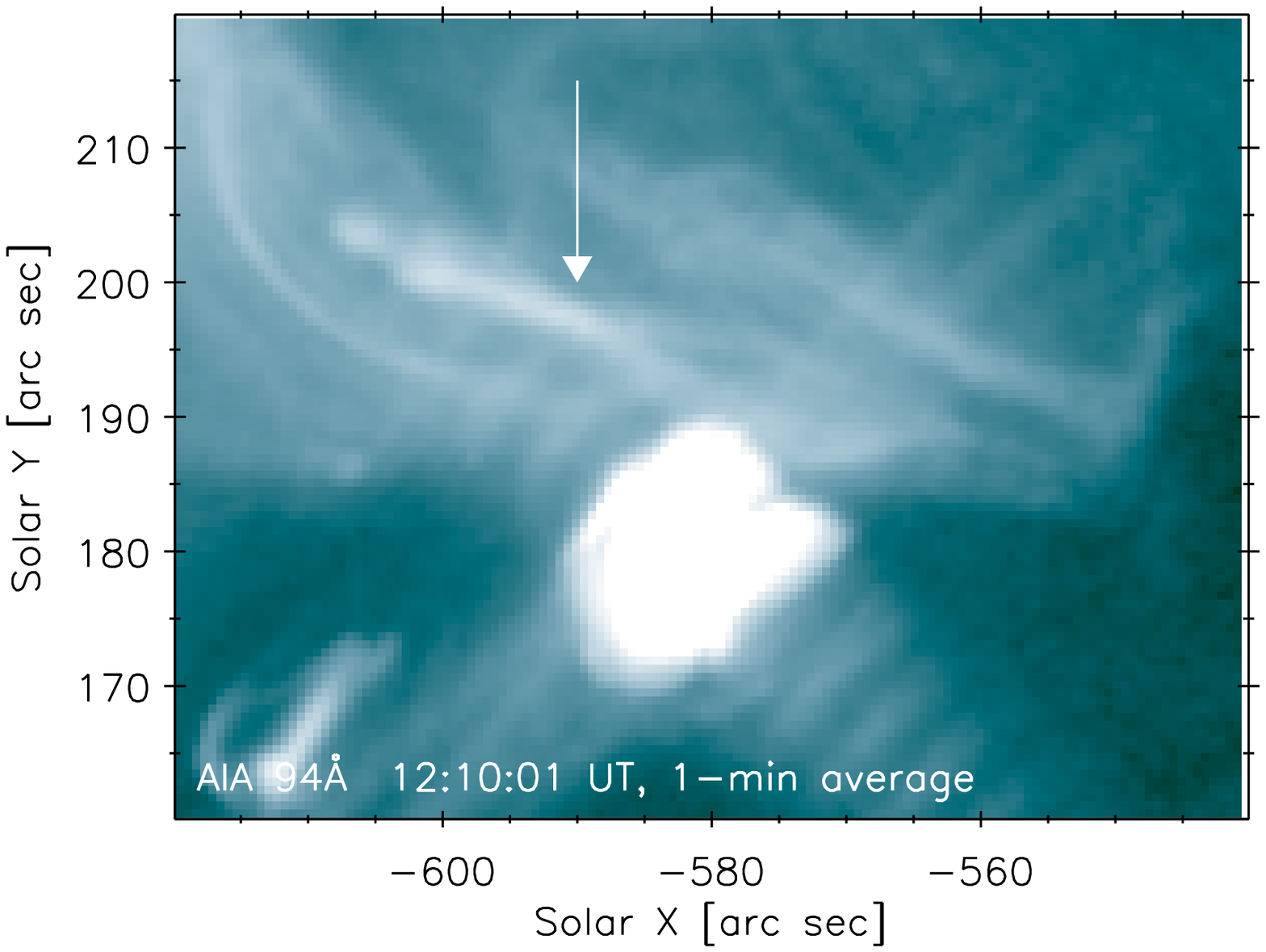}
        \includegraphics[width=6.45cm,clip,bb= 0  0 495 370]{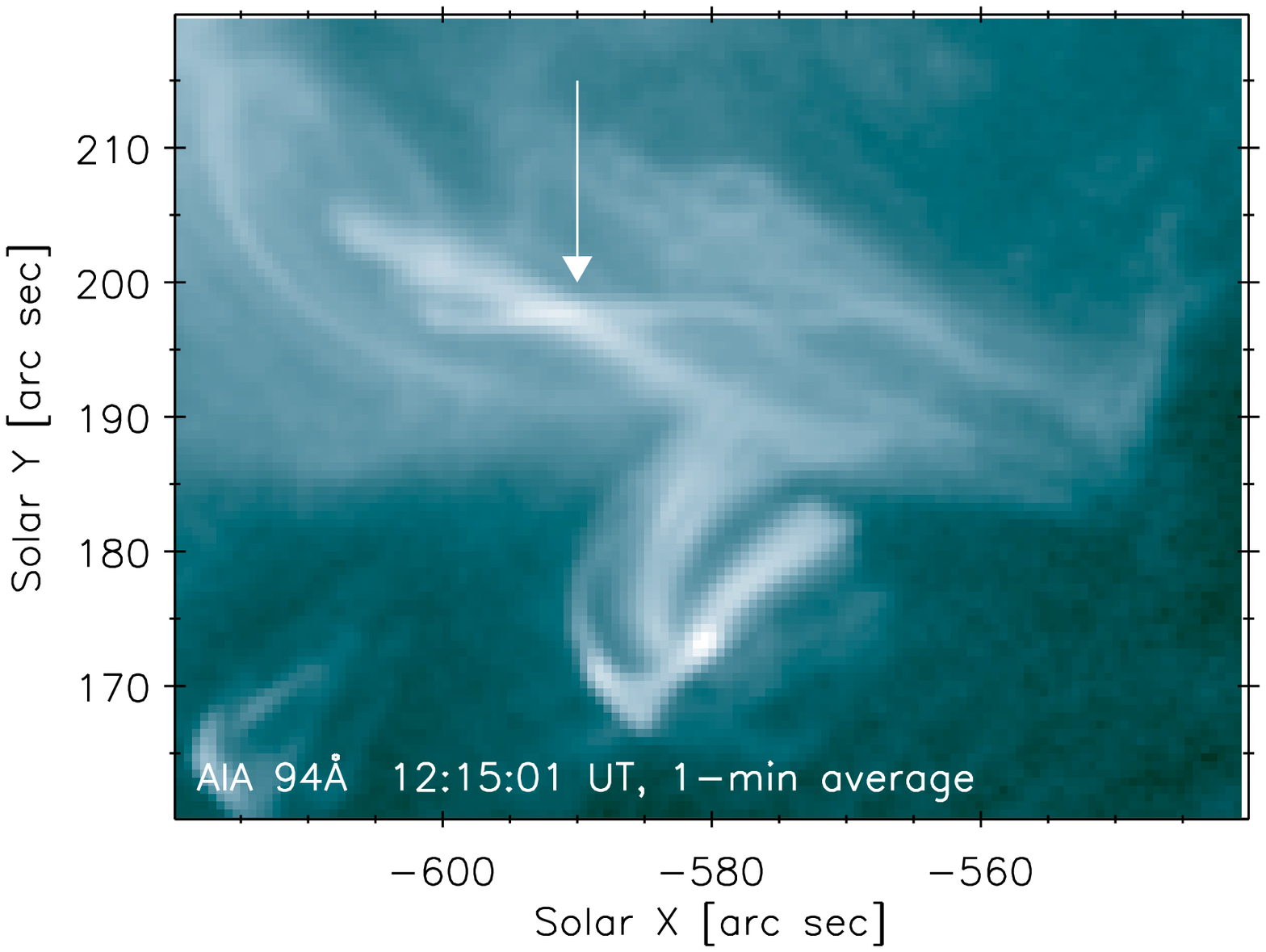}
        \includegraphics[width=5.58cm,clip,bb=67  0 495 370]{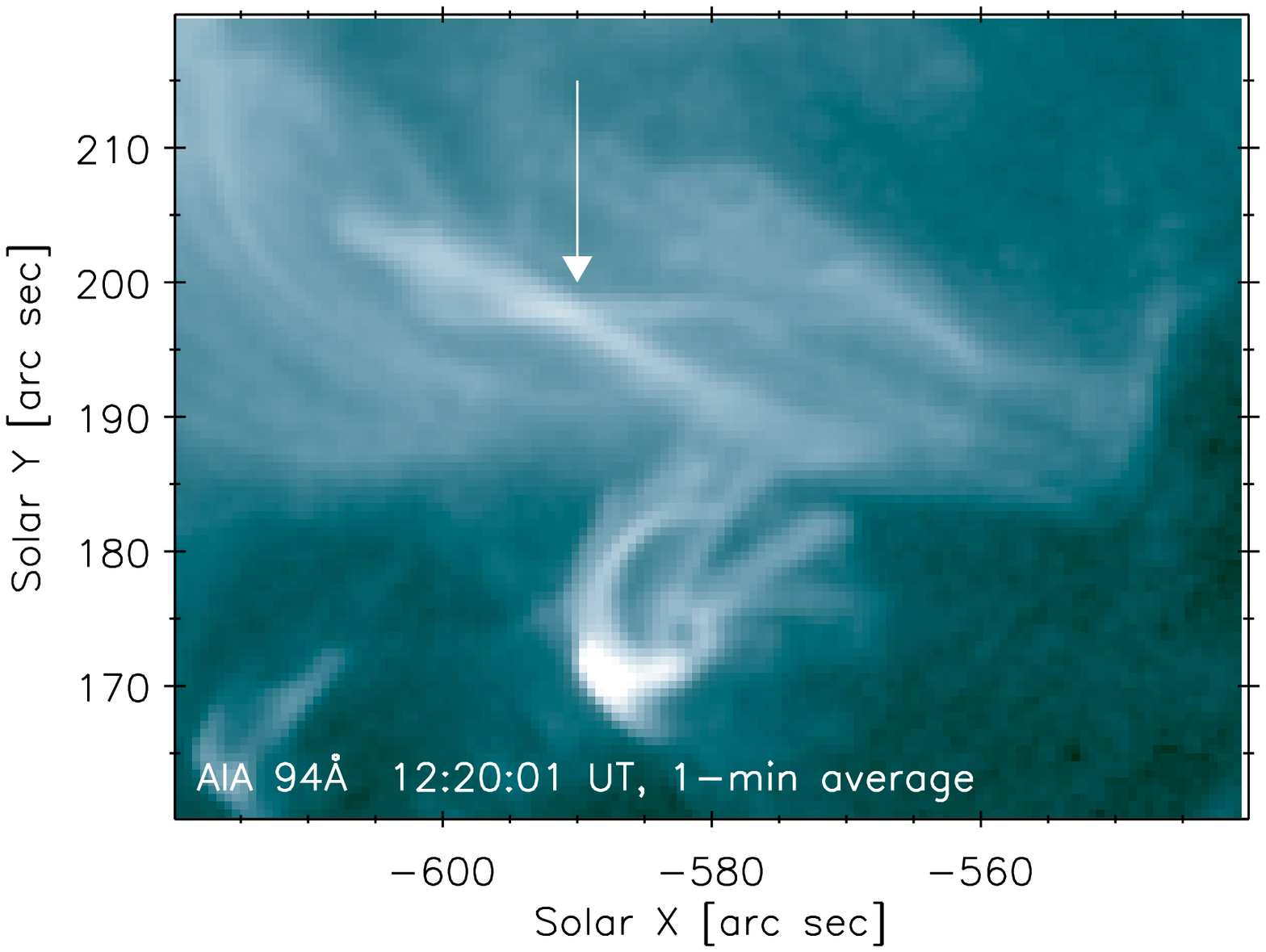}
        \includegraphics[width=5.58cm,clip,bb=67  0 495 370]{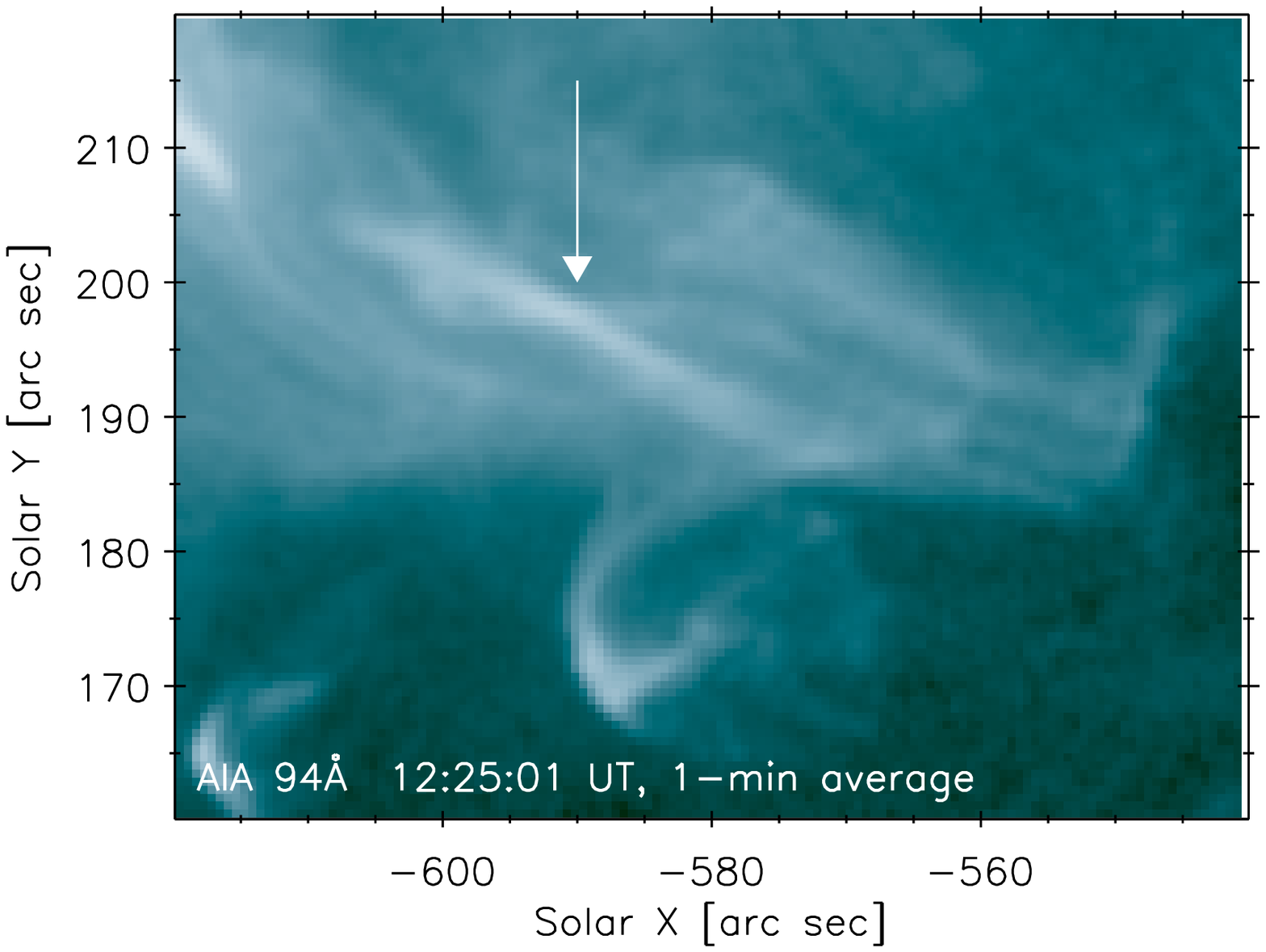}
        \caption{Evolution of the AIA 94\,\AA~emission after the GREGOR observations. The white arrow indicates the \ion{Fe}{XVIII} loops, whose footpoints lie in the ribbon. The ribbon itself is shown by the black contour in the \textit{top left} image. This contour corresponds to the 80 DN~s$^{-1}$\,px$^{-1}$ value in the IRIS SJI 1400\,\AA~observation at 12:00\,UT.}
       \label{Fig:Evolution_AIA94}
   \end{figure*}

\section{Interpretation and discussion}
\label{Sect:4}

Based on the behavior of knots 2 and 6, that is, on the oppositely oriented velocities along the ribbon, the similar behavior of the respective velocities, and the highly correlated intensity evolution, we propose that these two knots are footpoint locations of two flare loops that undergo slipping reconnection. It is known that the bright knots in the ribbon seen in the transition-region lines correspond well to footpoints of bright flare loops \citep[e.g.,][]{Dudik14}. Although small-scale bright knots are known to occur in the chromospheric ribbons \citep[e.g.,][]{Sharykin14}, our observations show that the slipping motion of these knots is also present in the chromospheric emission. This offers a direct indication that the energy release by slipping magnetic reconnection affects the deeper chromospheric layers where the \ion{Ca}{II} emission originates. In principle, the corresponding slipping reconnection \citep{Aulanier06} and the change in connectivity of the corresponding loops should be visible in EUV observations.

However, as a result of the order-of-magnitude difference in pixel size (0.026$\arcsec$ vs. 0.6$\arcsec$) and spatial resolution (0.1$\arcsec$ vs. 1.5$\arcsec$) of GREGOR and AIA, we are unable to resolve the corresponding individual thin flare loops in the AIA datasets. Nevertheless, the inspection of AIA images shows that post-reconnection loops anchored in the ribbon observed by GREGOR are indeed present and increase in brightness (Fig.~\ref{Fig:Evolution_AIA94}). These loops are prominent in AIA 94\,\AA, meaning that the temperature is sufficient to form \ion{Fe}{XVIII}. To confirm this, we also inspected the AIA 131\,\AA~and 171\,\AA~images (not shown). The loop emission is present but faint in 131\,\AA~and not visible in 171\,\AA, suggesting that they are indeed \ion{Fe}{XVIII} loops with a faint \ion{Fe}{XXI} component, that is, they are flare loops and not coronal loops emitting in \ion{Fe}{VIII}--\ion{Fe}{X}  \citep[cf.][]{ODwyer10,DelZanna13}.

The evolution of the AIA 94\,\AA~emission shows that the reconnection-heated loops indeed change their footpoints. At 12:15\,UT, two crossing loops anchored in the ribbon observed previously by GREGOR are clearly visible. The crossing point is indicated by the white arrow in Fig.~\ref{Fig:Evolution_AIA94}, while the location of the ribbon is plotted with a black contour in the 12:00\,UT image. These observations suggest that slipping reconnection may be continuing for about 10--15 minutes after the end of the GREGOR observations.

The AIA 94\,\AA~ also shows strong brightenings in the nearby circular ribbon during the C3.9 flare. This circular ribbon, whose appearance is probably the result of a magnetic null-point \citep[e.g.,][]{Masson09,Su13}, does not appear to be magnetically connected to the ribbon observed by GREGOR. Therefore, it is unlikely that the GREGOR ribbon is related to the topological discontinuities of the null-point. The change of connectivity in the AIA 94\,\AA~loops suggests that this ribbon is instead a portion of a QSL, in accordance with the presence of the slipping knots observed by GREGOR. However, owing to the nature of the optically thin emission, we cannot rule out that some of the loops located near the null-point are rooted in the same conjugate negative-polarity ribbon as the loops indicated by the white arrow (Fig.~\ref{Fig:Evolution_AIA94}) because the emission of both loop systems is overlaid at [$-575\arcsec$, 190$\arcsec$]. It is therefore possible that the two flux systems share a common QSL and that the flaring activity is physically related. However, we did not perform a topological analysis of the magnetic field. The active region 12205 is too complex, with multiple flux systems clearly visible in different AIA filters (Fig.~\ref{Fig:Overview}), high shear near the magnetic neutral line, and multiple flares. Consequently, the extrapolated magnetic field may not reflect the true magnetic configuration \citep{DeRosa09}, which is a result of the complex evolution of the flaring region \citep[see, e.g.,][]{Inoue14a,Inoue14b}.

%
\section{Concluding remarks}
\label{Sect:5}

We presented GREGOR observations of slipping bright knots in a small ribbon-like structure within a sunspot umbra. Two knots moving in the opposite direction along the ribbon exhibited highly correlated intensity changes. These observations indicate that the process of slipping magnetic reconnection has a direct chromospheric counterpart in the moving knots. Multiwavelength context observations were used to establish the relation of this ribbon to other flaring activity within AR 12205. We found that the slipping reconnection also occurs outside peaks of major flaring events.

The high spatial resolution of GREGOR (0.1$\arcsec$) shows that the slipping reconnection can occur on small spatial scales that are unresolvable by instruments such as AIA \citep[cf.][]{Dudik14}. Furthermore, the fast effective cadence (1\,s) of GREGOR allows detecting highly correlated intensity changes that would be difficult to discern at cadences an order of magnitude lower, as they are typical for AIA and IRIS \citep[e.g.,][]{Lemen12,Polito16}. The
reason is that the lifetimes of the knots can be of about one~minute, so that only a few snapshots would be observed by AIA or IRIS. Our observations illustrate the strong need for sub-arcsecond spatial resolution and fast cadences  so that the dynamics and the energy release in the solar atmosphere can be elucidated.

\begin{acknowledgements}
This work was supported by the grants 14-04338S (MS, JJ), P209/12/1652 (JD) of the Czech Science Foundation, and DE 7873.1 (CD, REL) of the Deutsche Forschungsgemeinschaft. WL was supported by the European Comission's FP-7 Collaborative Project No. 606862 ``F-CHROMA''. The support of the FP-7 Capacities Project No. 312495 ``SOLARNET'' and the  institutional support RVO:67985815 of the Academy of Sciences of the Czech Republic is also acknowledged. The 1.5 meter GREGOR solar telescope was built by a German consortium under the leadership of the Kiepenheuer Institute for Solar Physics in Freiburg with the Leibniz Institute for Astrophysics Potsdam, the Institute of Astrophysics G\"ottingen, and the Max Planck Institute for Solar System Research in G\"ottingen as partners, and with contributions by the Instituto de Astrof\'isica de Canarias and the Astronomical Institute of the Academy of Sciences of the Czech Republic. The SDO/AIA data are available by courtesy of NASA/SDO and the AIA science team. Hinode is a Japanese mission developed and launched by ISAS/JAXA, with NAOJ as domestic partner and NASA and STFC (UK) as international partners. It is operated by these agencies in co-operation with ESA and NSC (Norway). IRIS is a NASA small explorer mission developed and operated by LMSAL with mission operations executed at NASA Ames Research Center and major contributions to downlink funded by NSC, Norway, through an ESA PRODEX contract.
\end{acknowledgements}

\bibliographystyle{aa}
\bibliography{Slip_GREGOR}
%
%
\end{document}